\documentclass[pra,aps,amsmath,amssymb,twocolumn,showpacs]{revtex4}
\usepackage{graphicx}

\newcommand{\aver}[1]{\left< #1 \right>}
\newcommand{\ket}[1]{\left| #1 \right>}
\newcommand{\bra}[1]{{\left< #1 \right|}}
\DeclareMathOperator{\Tr}{Tr}
\DeclareMathOperator{\Rre}{Re}
\DeclareMathOperator{\Iim}{Im}

\begin{document}
\title{Entanglement, fidelity, and quantum-classical correlations 
with an atom walking in a quantized cavity field}
\author{S.V.~Prants, M.Yu.~Uleysky, and V.Yu.~Argonov}
\affiliation{Laboratory of Nonlinear Dynamical Systems,\\
V.I. Il'ichev Pacific Oceanological Institute of the Russian Academy of
Sciences,\\
690041 Vladivostok, Russia}
\date{\today}
\begin{abstract}
Stability and instability of quantum evolution are studied
in the interaction between a two-level atom with photon
recoil and a quantized field mode in an ideal cavity,
the basic model of cavity quantum electrodynamics (QED).
It is shown that the Jaynes-Cummings dynamics can
be unstable in the regime of chaotic walking of
the atomic center-of-mass in the quantized field of a standing wave
in the absence of any kind of interaction with
environment. This kind of quantum instability
manifests itself in strong variations of reduced
quantum purity and entropy, correlating with the
respective classical Lyapunov exponent, and in
exponential sensitivity of fidelity of quantum
states to small variations in the atom-field
detuning.
The connection between quantum entanglement and fidelity and
the center-of-mass motion is clarified
analytically and numerically for a few regimes of that motion.
The results are illustrated with two specific
initial field states: the Fock and coherent ones.
Numerical experiments demonstrate various manifestations
of the quantum-classical correspondence, including
dynamical chaos and fractals, which can be, in
principle, observed in real experiments with atoms
and photons in high finesse cavities.

\end{abstract}
\maketitle

\section{Introduction}

Entanglement is a birth-mark of quantum mechanics.
A pure state of a closed quantum system evolves as a pure
state, and all the measures of entanglement remain
constant during the evolution. However, if the system is
composed of interacting subsystems, the reduced density
matrix of one of them~--- a measure of coherence
lost in this subsystem~--- does not evolve in a unitary way.
That is the reason of the loss of coherence in a
system interacting with an environment
composed of a large number of degrees of freedom.
A lot of work has been done to understanding
decoherence, dynamics of entanglement, and quantum
chaos. One of the themes is the behavior of entanglement
in bipartite quantum systems in the
parameter regimes where their underlying classical
analogues are chaotic in the sense of exponential
sensitivity to small changes in initial conditions.

The basic models of quantum optics and cavity quantum electrodynamics
are the Jaynes-Cummings model \cite{JC}, describing the interaction between
a two-level atom at rest and a quantized mode of the radiation field, and
the Dicke model \cite{D},
describing the interaction of a collection of two-level atoms,
located within a distance much smaller than the radiation wavelength, 
with a quantized mode of the radiation field.
It is known for a long time \cite{BTZ, AMS}
that the classical limit of the Dicke model
can be chaotic only beyond the rotating-wave approximation.
It has been found that initial increase in the reduced linear entropy 
is faster for the initial states, prepared as wave
packets centered at chaotic regions of the classical
phase space, as compared to regular ones \cite{F98, XH04}. The similar property has been found
in the model of weakly-coupled quantum tops \cite{MS99} and
in the model of two-component Bose-Einstein
condensates \cite{XH05}.

In the above cited and other similar works,
closed quantum systems have been considered,
and the connection has been studied between
quantum entanglement and dynamical chaos in
the classical version of a quantum system
under consideration, i.~e.~between two
{\it different} systems.
However, the classical limit of a quantum system is not unique. For
example, there are different classical limits of the Jaynes-Cummings
model \cite{JC} and its generalization, the Dicke model \cite{D},
which even have different names: neoclassical, quasiclassical, and
semiclassical \cite{P97, B04}. One may treat atoms quantum mechanically and
the field classically. There is a choice how to couple the
atoms with their radiation field. The classical limit
can be taken for both the atomic and field subsystems.
As a result, one derives different equations of motion
for expectation values of quantum operators for the same physical system
whose phase spaces may have different properties.
It is a problem to decide which classical limit is
``the best one'' and should be used for comparison with
quantum treatment of the system. 

In this paper we study the
connection between the purely quantum characteristics, such as entanglement 
and fidelity, and classical chaos in {\it the same} system
with dynamically coupled quantum and classical degrees of freedom.
The {\it quantum dynamics} of a system with {\it classical
degrees of freedom} is of great interest not only from the point of view 
of quantum chaos but from the view point of the correspondence between
quantum and classical mechanics as well.

The main purpose of this paper is to show that entanglement between the atomic
and field quantum states correlates with the center-of-mass motion of an atom
in the standing-wave cavity field. Regular center-of-mass motion corresponds
to a regular evolution of the reduced quantum entropy and  fidelity of quantum
states, whereas quantum evolution is unstable under a chaotic (random) walking
of the atom. Quantum instability may arise in quantum-classical hybrid 
without introducing a bath with infinitely many
degrees of freedom or any kind of noise to model 
environment whose role is played in our model by a translational degree of freedom. We prove that  correlations 
between purely quantum characteristics, such as entanglement and fidelity 
of the atom-field states, and the classical measure of motion~---
the maximal Lyapunov exponent~--- arise in a natural way with 
a two-level atom moving in a single-mode quantized cavity field. By numerical
simulation we demonstrate various manifestations of the quantum-classical
correspondence, including dynamical chaos and dynamical fractals, that may be, in principle, found
in real experiments with atoms and photons in high finesse cavities. 

Atoms, interacting with the electromagnetic field, change both their 
internal electronic states and external translational states, the process 
known as a photon recoil. In the strong-coupling regime, the respective 
Hamiltonian describes the interaction between the field, electronic, and 
translational degrees of freedom. In our previous papers \cite{PS01,P02,AP03} 
we have studied 
the Hamiltonian dynamics of a two-level atom in a self-consistent 
standing-wave cavity field in the semiclassical approximation in the 
limit of an infinite number of photons and found  both regular and chaotic 
regimes of motion of the atomic center of mass. The chaotic walking of 
atoms takes place in a stationary standing wave and is quantified 
by the positive values of the maximal Lyapunov exponent \cite{PS01}. Typical  
chaotic atomic trajectories consist of intervals of random walking interrupted 
by regular oscillations near the bottom of some wells of the optical 
potential and long ballistic flights with practically constant velocity 
(so-called L\'evy flights \cite{PEZ02}). In the language of dynamical system's theory, 
there is a fractal-like chaotic invariant set consisting of all chaotic and unstable 
periodic orbits in the phase space of the strongly coupled atom-field system. 
As a sequence, a scattering of atoms even  at a 1D standing electromagnetic 
wave is fractal with prominent self-similarity of their scattering function, 
for example, in dependence of exit time on the atomic initial momentum 
\cite{AP03}.   

Classical instability is usually defined as an exponential separation of two
nearly trajectories in time with an asymptotic rate given by the maximal Lyapunov
exponent $\lambda$. Perfectly isolated quantum systems are unitary and cannot 
be unstable
in this sense even if their classical limits are chaotic \cite{C91}. 
It has been long ago proposed \cite{Peres} to measure quantum instability 
by the decay of fidelity or overlap 
of two initially identical wave functions that evolve under
slightly different Hamiltonians. In a number of numerical studies (see, for 
example Refs.~\cite{Caves,JP01,JSB01,BC02,PSZ03} and cited therein) for a 
variety of classically chaotic models, it has been found that the overlap 
decay is, in general, not universal and depends on the strength of 
perturbations in Hamiltonians and other factors determining which decay  regime 
prevails, algebraic, Gaussian, or exponential one. The usual strategy is to 
compare the fidelity decay to a separation of nearby trajectories in the 
phase space of a classical analogue of the quantum system under consideration. 
With our dynamically-coupled quantum-classical hybrid, a correlation 
between quantum and classical measures of instability arises in a natural way 
as a result of coupling between different characteristics of the same system.

The paper is organized as follows. In Sec.~\ref{section_main_eqs}
we derive the Hamilton-Schr\"odinger nonlinear equations, describing 
interaction between  
a two-level atom with the classical translational degree of freedom and 
a single-mode quantized field in an ideal cavity, and discuss possible 
regimes of the center-of-mass motion in dependence on the values 
of the detuning of the atom-field resonance.  We express linear and 
von Neumann entropies (measures of the atom-field entanglement) and 
fidelity  (measure of instability of the Jaynes-Cummings quantum evolution)  
in terms of the respective probability amplitudes. In some cases 
we find exact and approximate solutions of the equations of motion.  
In Sec.~\ref{section_fock} we study correlations between  
the entanglement and fidelity and the mode of the center-of-mass motion  
in a Fock-state cavity field. In this case the infinite-dimensional set 
of the equations of motion reduces to a few coupled nonlinear equations 
which are analyzed for different regimes of the center-of-mass motion.  
In fact, we deal with entanglement between  two similar quantum 
systems, a two-level atom and a three- (or two-) level field.
We find exact and approximate solutions for the entanglement 
in the Fock field and demonstrate numerically the close correspondence 
between these quantum characteristics  and underlying classical chaos.  
In Sec.~\ref{section_coherent} we report on correlations found 
between the entanglement and fidelity and classical random walking 
of an atom in a coherent-state field which are  quantified in terms
of the respective classical maximal Lyapunov exponent. Section \ref{section_concl} is for concluding remarks.   

\section{\label{section_main_eqs}A two-level atom walking in a quantized cavity field}
\subsection{Hamilton-Schr\"odinger equations of motion}

To specify the problem we consider the standard model in cavity QED with the 
Jaynes-Cummings Hamiltonian~\cite{JC}
\begin{multline}
\hat H=\frac{\hat p^2}{2m_a}+\frac{1}{2}\hbar\omega_a\hat\sigma_z+\hbar\omega_f\left(\hat a^\dag\hat a+\frac{1}{2}\right)-\\
-\hbar\Omega_0\left(\hat a^\dag\hat\sigma_-+\hat a\hat\sigma_+\right)\cos{k_f\hat x},
\label{Janes-Cum}
\end{multline}
which describes the interaction between a two-level atom (with lower $\ket{1}$ and
upper $\ket{2}$ states, the transition frequency $\omega_a$, and the Pauli operators $\hat\sigma_{\pm,z}$)
and a quantized electromagnetic-field mode (with creation $\hat a^\dag$ and 
annihilation
$\hat a$ operators) forming a standing wave with the frequency
$\omega_f$ and the wave vector $k_f$ in an ideal cavity. 
The atom and field become dynamically entangled by
their interaction with the state of the combined system after the interaction
time $t$
\begin{equation}
\ket{\Psi(t)}=\sum_{n=0}^\infty\left(a_n(t)\ket{2,n}+b_n(t)\ket{1,n}\vphantom{\sqrt{M}}\right)
\label{Psi}
\end{equation}
to be expanded over the Fock field states $\ket{n}$, $n=0,1,\dots$. Here 
$a_n(t)$ and $b_n(t)$ are the
complex-valued probability amplitudes to find the field in the state $\ket{n}$ and the
atom in the states $\ket{2}$ and $\ket{1}$, respectively. In the process of emitting and
absorbing photons, atoms not only change their internal electronic states
but their external translational states change as well due to the photon
recoil effect. If atoms are not too cold and their average momenta are large
as compared to the photon momentum $\hbar k_f$, one can describe the translational degree
of freedom classically. 

The whole dynamics is now governed by the Hamilton-Schr\"odinger
equations \cite{PU03} that have the following normalized form in the frame rotating
with the frequency $\omega_f(n+1/2)$:
\begin{equation}
\begin{gathered}
\dot x=\omega_r p,\\
\dot p=-2\sin x\sum_{n=0}^\infty\sqrt{n+1}\Rre{\left(a_nb_{n+1}^*\right)},\\
i\dot a_n=-\frac{\delta}{2}a_n-\sqrt{n+1}\,b_{n+1}\cos x,\\
i\dot b_{n+1}=\frac{\delta}{2}b_{n+1}-\sqrt{n+1}\,a_n\cos x
\end{gathered}
\label{mainsys}
\end{equation}
with simple solution for the probability amplitude $b_0(\tau)$
\begin{equation}
b_0(\tau)=b_0(0)e^{-i\delta \tau/2}. 
\label{b0treal}
\end{equation}
Here $x\equiv k_f\aver{\hat x}$ and $p\equiv\aver{\hat p}/\hbar k_f$ are the atomic center-of-mass position and momentum, respectively.
Dot denotes differentiation with respect to dimensionless time $\tau\equiv\Omega_0 t$, where
$\Omega_0$ is the amplitude coupling constant. The normalized recoil frequency 
$\omega_r\equiv\hbar k_f^2/m_a\Omega_0\ll 1$ 
and the atom-field detuning $\delta\equiv(\omega_f-\omega_a)/\Omega_0$ are the control parameters.
This set possesses an infinite number of the integrals of motion
\begin{equation}
R_n\equiv |a_n|^2+|b_{n+1}|^2=\text{const},
\quad \sum_{n=0}^\infty R_n\leqslant 1,
\label{Rn}
\end{equation}
and conserves the total energy
\begin{multline}
E\equiv\frac{\omega_r  p^2}{2}-\frac{\delta}{2}\sum_{n=0}^\infty\left(|a_n|^2-|b_{n+1}|^2\right)-\\
-2\cos x\sum_{n=0}^\infty\sqrt{n+1}\Rre{\left(a_nb_{n+1}^*\right)}.
\label{Energy}
\end{multline}

By introducing new variables 
\begin{equation}
\begin{aligned}
&u_n\equiv 2\Rre{\left(a_nb_{n+1}^*\right)},\\
&v_n\equiv -2\Iim{\left(a_nb_{n+1}^*\right)},\\
&z_n\equiv\left|a_n\right|^2-\left|b_{n+1}\right|^2,
\end{aligned}
\label{uvz_def}
\end{equation}
we can rewrite the set (\ref{mainsys}) in the following form: 
\begin{equation}
\begin{gathered}
\dot x=\omega_r p,\qquad \dot p=-\sin{x}\sum\limits_{n}\sqrt{n+1}\,u_n,\\
\begin{aligned}
\dot u_n&=\delta v_n,\\
\dot v_n&=-\delta u_n+2\sqrt{n+1}\,z_n\cos{x},\\
\dot z_n&=-2\sqrt{n+1}\,v_n\cos{x}  
\end{aligned}
\end{gathered}
\label{UVZSystem}
\end{equation}
with the respective integrals of motion 
\begin{equation}
u_n^2(\tau)+v_n^2(\tau)+z_n^2(\tau)=R_n^2=\text{const} 
\label{uvz_conserv}
\end{equation}
and the total energy
\begin{equation}
E=\frac{\omega_r p^2}{2}-\frac{\delta}{2}\sum\limits_{n=0}^{\infty} 
z_n-\cos{x}\sum\limits_{n=0}^{\infty}\sqrt{n+1}\,u_n.
\label{UVZEnergy}
\end{equation}
The inverse transformation gives us information about the moduli of 
the probability amplitudes only
\begin{equation}
\left|a_n\right|^2=\frac{R_n+z_n}{2},\qquad
\left|b_{n+1}\right|^2=\frac{R_n-z_n}{2}.
\label{a2b2}
\end{equation}

The Hamilton-Schr\"odinger equations (\ref{mainsys}) and (\ref{UVZSystem}) describe
a quantum-classical hybrid with the classical part
(the first two equations in the sets), driven by the
quantum probability amplitudes, and the quantum
one (the other equations in the sets) driven by
the envelope of the standing wave. The quantum
part is still unitary and therefore cannot suffer
from decoherence induced by the classical part. In
other words, the total quantum entropy is constant
during the evolution. However, the quantum part is
a system composed of two quantum subsystems
each of which may be characterized by a reduced quantum entropy. 
The role of the classical
motion for the reduced entropy is not trivial
and will be clarified in the other sections.

\subsection{Entanglement and fidelity of the atom-field states}

The atomic population inversion $z(\tau)$ is a difference of probabilities to 
find an atom at the moment of time $\tau$ in the excited and ground states 
\begin{equation}
z\equiv\sum\limits_{n=0}^{\infty}|a_n|^2-\sum\limits_{n=0}^{\infty}|b_n|^2=\sum\limits_{n=0}^{\infty}z_n-|b_0|^2.
\label{to_z}
\end{equation}
It is an important characteristic that can be measured in experiments. 
The entanglement between the internal atomic and field degrees of freedom can
be characterized by the quantity known as purity 
\begin{equation}
P(\tau)\equiv\Tr\rho_a^2(\tau),
\label{Puritydef}
\end{equation}
where $\rho_a(\tau)$ is the reduced atomic density matrix
\begin{equation}
\rho_a(\tau)\equiv\sum_{n=0}^\infty\bra{n}\,\rho(\tau)\ket{n}
\label{Matrix}
\end{equation}
with the total density matrix to be $\rho(\tau)\equiv\ket{\Psi(\tau)}\bra
{\Psi(\tau)}$. Purity is maximal if an atom is in
one of its energetic states $\ket{1}$ or $\ket{2}$, i.~e. $P_\text{max}=
\Tr\rho_a^2=\Tr\rho_a=1$.
Purity is minimal if $\rho_a=I/2$, i.~e. $P_\text{min}=1/2$, where $I$ 
is the identity matrix. In terms
of the probability amplitudes, it is given by
\begin{equation}
\begin{gathered}
P=A^2+B^2+2|C|^2,\\
A\equiv\sum\limits_{n=0}^{\infty}|a_n|^2,\quad B\equiv\sum\limits_{n=0}^{\infty}|b_n|^2,\quad   C\equiv\sum\limits_{n=0}^{\infty}a_nb_n^*.
\end{gathered}
\label{Purity}
\end{equation}
As similar standard measures of quantum ``disorder'', we will use 
the reduced linear entropy
\begin{equation}
S_L\equiv 1-\Tr\rho_a^2=1-P
\label{to_S_L}
\end{equation}
and the reduced von Neumann entropy 
\begin{equation}
\begin{gathered}
S_N\equiv-\Tr(\rho_a\ln\rho_a)=-\lambda_1\ln\lambda_1-\lambda_2\ln\lambda_2,\\
\lambda_{1,2}=\frac{1}{2}\pm\sqrt{\frac{1}{4}+|C|^2-AB}, 
\end{gathered}
\label{to_S_N}
\end{equation}
where $\lambda_{1,2}$ are eigenvalues of the matrix $\rho_a$.

To quantify instability of the Jaynes-Cummings quantum evolution 
we use the fidelity $f(\tau)$ which is the overlap of two states $\ket{\Psi_1(\tau)}$ and
$\ket{\Psi_2(\tau)}$,
identical at $\tau=0$, that evolve under two Hamiltonians (\ref{Janes-Cum}) 
with slightly different values of one of the system's parameters $\kappa$.  
In terms of the probability amplitudes it has the form 
\begin{multline}
f(\tau,\,\kappa,\Delta\kappa)\equiv\left|\left<\Psi(\tau,\kappa)|\Psi(\tau,\kappa+\Delta\kappa)\right>\right|^2=\\
=\left|\sum\limits_{n=0}^{\infty}\left(\vphantom{\sqrt{M}}
a_n^*(\tau,\kappa)a_n(\tau,\kappa+\Delta\kappa)+
\right.\right.\\
\left.\vphantom{\sum\limits_{n=0}^{\infty}}\left.\vphantom{\sqrt{M}}
+b_n^*(\tau,\kappa)b_n(\tau,\kappa+\Delta\kappa)
\right)\right|^2,
\label{to_fidelity}
\end{multline}
where $\Delta\kappa$~is a small variation of the parameter $\kappa$.
In difference from the reduced quantum
entropies introduced above, fidelity is a
characteristic of the whole quantum system,
not of its part.

\subsection{\label{subsection_analyt}Analytical solutions 
with arbitrary field states}

The Jaynes-Cummings dynamics of a single two-level atom at rest,  
interacting with a single mode of the quantized field, has been 
studied in detail by many authors (for a review, see~\cite{SK93}). 
We are interested here how the dynamical coupling between the Jaynes-Cummings  
and classical degrees of freedom changes both the quantum and classical 
dynamics.              

The Hamilton-Schr\"odinger equations (\ref{mainsys}) are, in general, 
non-integrable. The
type of the center-of-mass motion depends strongly on the values of 
the detuning
$\delta$. In the limit of zero detuning and with initially excited or deexcited
atoms, the optical potential disappears, and atoms moves with a constant
velocity $\dot x=\omega_r  p_0$. The quantum evolution is periodic with the 
period $\pi/\omega_r  p_0$, and exact
solutions for purity, von Neumann entropy,  and other 
quantum
characteristics can be found in the explicit form. 
With arbitrary initial field and atomic states,
the variables $u_n$ are constants, $u_n=u_n(0)$, for each $n$. The Hamilton
equations for the translational degree of freedom
are closed, and their solution can be easily
found in terms of the Jacobi elliptic functions
\cite{PS01}. For each $n$, we get the following exact solution:
\begin{equation}
\begin{gathered}
\begin{aligned}
&z_n(\delta=0)=\\
&=\mp\sqrt{R_n^2-u_n^2(0)}\sin\left(2\sqrt{n+1}\int\cos x(\tau)\,d\tau+\varphi_n\right),
\end{aligned}\\
\varphi_n\equiv\mp\arcsin\left(z_n(0)/\sqrt{R_n^2-u_n^2(0)}\right),
\end{gathered}
\label{Res}
\end{equation}
that describes a frequency modulated signal.
Exact solutions for the amplitudes $a_n$ and $b_n$ are the following:
\begin{multline}
a_n(\delta=0)=a_n(0)\cos\left(\sqrt{n+1}\int\cos x(\tau)\,d\tau\right)+\\
+ib_{n+1}(0)\sin\left(\sqrt{n+1}\int\cos x(\tau)\,d\tau\right),
\end{multline}
\begin{multline}
b_{n+1}(\delta=0)=b_{n+1}(0)\cos\left(\sqrt{n+1}\int\cos x(\tau)\,d\tau\right)+\\
+ia_n(0)\sin\left(\sqrt{n+1}\int\cos x(\tau)\,d\tau\right),
\end{multline}
where $a_n(0)$ and $b_{n+1}(0)$ are the initial complex values of 
$a_n$ and $b_{n+1}$.
In the Raman-Nath limit, $p\simeq p_0=\text{const}$, the solution 
(\ref{Res}) is simplified
\begin{multline}
z_n(\delta=0)\simeq\\
\simeq\mp\sqrt{R_n^2-u_n^2(0)}\sin\left(\frac{2\sqrt{n+1}}{\omega_r p_0}\sin\omega_r p_0\tau+\varphi_n\right).
\label{RN}
\end{multline}

With the detuning being large, $|\delta|\gg 1$, the optical potential is 
shallow, atom
moves with almost a constant velocity, $\simeq\omega_r  p_0$, slightly 
modulated by the standing
wave, and its inversion oscillates with a small depth (excepting for the case
of the so-called Doppler-Rabi resonance with maximal Rabi oscillations that
occur at the condition $|\delta|\simeq\omega_r|p_0|$ \cite{UKP03}). If the atomic 
kinetic energy $\omega_r  p^2/2$ 
is not enough to overcome barriers of the optical potential, the atomic center
of mass oscillates nonlinearly in one of the potential wells.

With very fast atoms, $\omega_r|p_0|\gg\max(|\delta|, 2\sqrt{n+1})$, 
or large detunings,
$|\delta|\gg\max(\omega_r|p_0|, 2\sqrt{n+1})$, one can get
the approximate  amplitude-modulated solutions for $z_n$
\begin{equation}
\begin{gathered}
z_n\simeq z_n(0)-\frac{2\sqrt{(R_n^2-z_n^2(0))(n+1)}}{\omega_r p_0}
\cos(\delta\tau+\phi_n)\sin x,\\
\omega_r|p_0|\gg\max(\delta, 2\sqrt{n+1});\\
\begin{aligned}
z_n\simeq z_n(0)&+\frac{2\sqrt{n+1}u_n(0)}{\delta}-\\
-&\frac{2\sqrt{(R_n^2-z_n^2(0))(n+1)}}{\delta}
\cos x\sin(\delta\tau+\phi_n),
\end{aligned}\\
\delta\gg\max(\omega_r|p_0|, 2\sqrt{n+1}),
\end{gathered}
\label{z2}
\end{equation}
where $\phi_n\equiv\arcsin\left(u_n(0)/\sqrt{R_n^2-z_n^2(0)}\right)$.
In both the cases, we can find  the respective approximate solutions for 
fidelity, i.e. overlap (\ref{to_fidelity}) of two initially identical 
quantum states, evolving under Hamiltonians with slightly different 
values of the detuning $\Delta\delta$ 
\begin{equation}
f(\tau,\,\delta,\Delta\delta) \simeq A^2(0)+B^2(0)+2A(0)B(0)\cos\Delta\delta\tau,
\end{equation}
where $A(0)$, $B(0)$ are initial values of $A$ and $B$ defined in 
(\ref{Purity}).

Inspecting the Hamilton-Schr\"odinger equations (\ref{mainsys}),  
we see that the translational motion is described by the equation for  
a nonlinear oscillator with the frequency modulation which is caused by 
the Jaynes-Cummings dynamics. It has been proven in Ref.~\cite{P02} for this type 
of equations that,  
as the result of interaction of nonlinear resonances, there arises 
a stochastic layer in the respective classical phase space. The width of this 
layer depends on the values of the detuning $\delta$. With the detunings 
of the order $|\delta|\lesssim 1$, the atomic center of mass can walk in an 
erratic way inside a cavity with a stationary standing wave. 
This type of motion may be
called a chaotic or random walking, and it is quantified by positive values
of the maximal Lyapunov exponent $\lambda$. For a classical standing wave,  
the Hamiltonian chaos has been studied in detail in Refs. \cite{PS01,PEZ02,AP03}
where the dynamical effects of L\'evy flights and atomic fractals have been 
found and described. 
    
\section{\label{section_fock}Entanglement, fidelity, and quantum-classical correlation 
in a Fock field}

In this section we describe a quantum-classical correlation
for a two-level atom moving in the quantized field prepared initially in a Fock state. If the
cavity mode is prepared in the state $\ket{n}$ with
an exactly specified number of photons $n$, the
infinite-dimensional set of the Hamilton-Schr\"odinger
equations (\ref{UVZSystem}) reduces to the following finite-dimensional
set which is written in terms of the Bloch-like variables:
\begin{equation}
\begin{gathered}
\dot x=\omega_r p,\qquad
\dot p=-\left(\sqrt{n}\,u_{n-1}+\sqrt{n+1}\,u_n\right)\sin x,\\
\dot u_{n-1}=\delta v_{n-1},\qquad
\dot u_{n}=\delta v_{n},\\
\dot v_{n-1}=-\delta u_{n-1}+2\sqrt{n}\,z_{n-1}\cos x,\quad\\
\dot v_{n}=-\delta u_{n}+2\sqrt{n+1}\,z_{n}\cos x,\\
\dot z_{n-1}=-2\sqrt{n}\,v_{n-1}\cos x,\qquad
\dot z_{n}=-2\sqrt{n+1}\,v_{n}\cos x.
\end{gathered}
\label{F}
\end{equation}
In fact, we deal with a two-level quantum system (an atom)
entangled with a three-level quantum system (a field
state which is a superposition of the three Fock states $\ket{n-1}$, $\ket{n}$, and $\ket{n+1}$). The atomic purity (\ref{Puritydef}) can be now
expressed in terms of the components of the
atomic population inversion $z(\tau)=z_{n-1}(\tau)+z_n(\tau)$ and
the integrals of motion $R_{n-1}$ and $R_n$ (\ref{Rn})
\begin{equation}
P_n=\frac{1}{2}\left[1+(z_n+z_{n-1})^2+(R_n+z_n)(R_{n-1}-z_{n-1})\right].
\end{equation}

With the atom prepared initially
in one of its energetic states, say, in the excited
state $\ket{2}$, and the field, prepared in a Fock state $\ket{n}$, 
we get the simplest possible kind 
of entanglement between these two two-level quantum
systems. The respective purity is extremely simplified
\begin{equation}
P_n=\frac{1}{2}\left(1+z_n^2\right).
\end{equation}
We can now analyze in detail the correlations between the quantum entanglement
and the center-of-mass motion.
At exact resonance $\delta=0$, atoms, initially prepared in one of the energetic states, fly through a cavity with a constant velocity $p_0$, and
using (\ref{Res}) we get exact solution for the purity
\begin{equation}
P_n(\delta=0)=\frac{1}{2}+\frac{1}{2}\cos^2\left(\frac{2\sqrt{n+1}}{\omega_r p_0}\sin\omega_r p_0\tau\right),
\label{zres}
\end{equation}
oscillating periodically between $1/2$ and $1$.

In the other limiting
cases of fast atoms, $\omega_r|p_0|\gg\max(|\delta|, 2\sqrt{n+1})$, and large detunings,
$|\delta|\gg\max(\omega_r|p_0|, 2\sqrt{n+1})$,
the atomic center-of-mass motion is regular
(nonlinear oscillations in a well of the optical potential
or a ballistic flight with almost a constant velocity
$\omega_r p_0$ slightly modulated by the standing
wave), the atomic inversion oscillates with
a small depth, and the atomic purity is $P_n\simeq 1$.
But oscillations of the purity can be
large with periodically maximal entanglement between
the atom and the Fock field, if the condition
of the Doppler-Rabi resonance, $\omega_r|p|\simeq |\delta|$, is
fulfilled. When the atom comes into the resonance
with one of the running waves, composing a
standing wave, the Rabi oscillations become
maximal. Using the solution for the atomic
inversion found in \cite{AP03}, it is easy to
get the respective atomic purity
\begin{multline}
P_n(|\delta|\simeq\omega_r|p|)\simeq\\
\simeq\frac{1}{2}+
\frac{1}{2}\left(\frac{\left(|\delta|-\omega_r|p_0|\right)^2}{\Omega_n^2}+
\frac{\sqrt{n+1}}{\Omega_n^2}\cos\Omega_n\tau\right)^2,
\label{DR}
\end{multline}
oscillating periodically with the Rabi frequency
$\Omega_n\equiv\sqrt{\left(|\delta|-\omega_r|p_0|\right)^2+n+1}$ and the maximal
amplitude at $|\delta|=\omega_r|p_0|$.

\begin{figure*}
\includegraphics[width=0.9\textwidth]{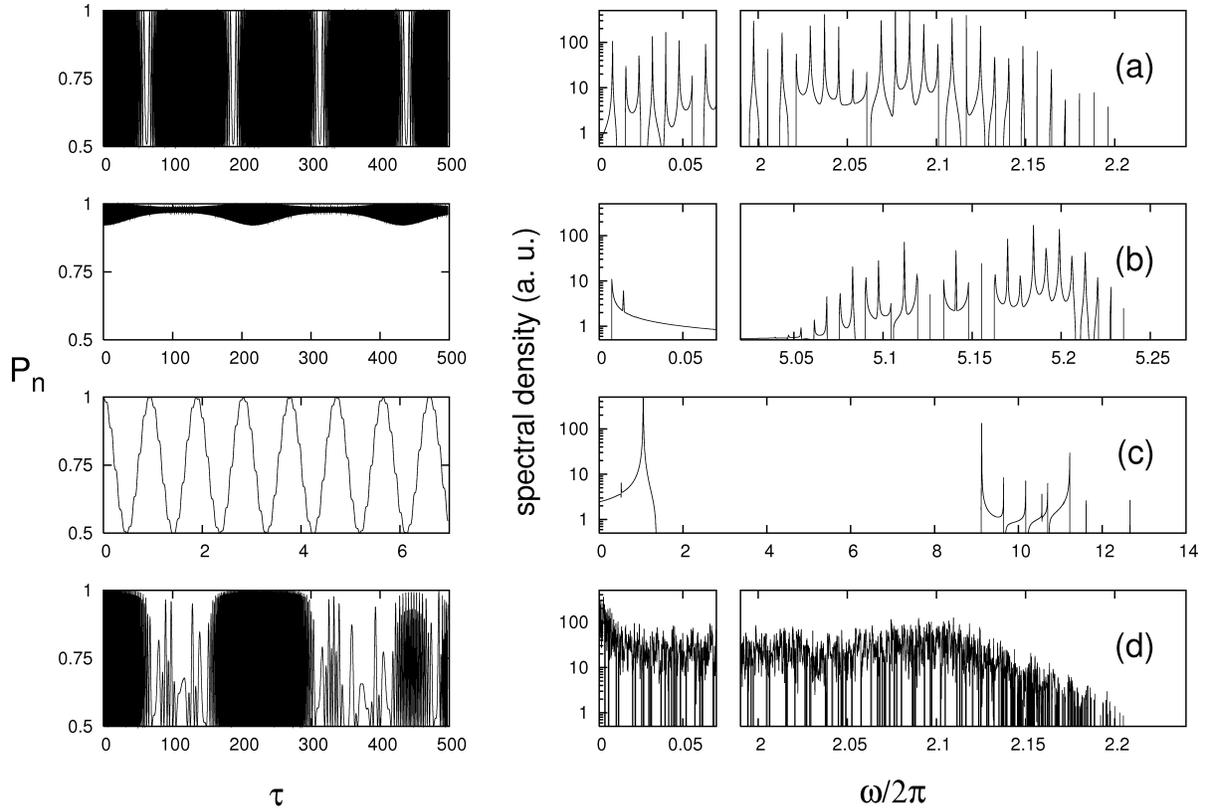}
\caption{Left panel: time evolution 
($\tau$ is in units of $\Omega_0^{-1}$) of the atomic purity $P_n$
with an atom initially prepared in the excited state and the field
initially in a Fock state with $n=10$. (a) Exact resonance,
$\delta=0$ and $p_0=25$, (b) a ballistic atom, $\delta=32$ and $p_0=25$, (c) the Doppler-Rabi
resonance with $\delta=32$ and $p_0=32000$, (d) chaotic walking
with $\delta=0.4$ and $p_0=25$. Right panel: the respective power spectra,
with the frequency $\omega$ in units of $\Omega_0$.}
\label{fig1}
\end{figure*}
Chaotic walking of the atomic center of mass in
a standing wave which is initially prepared in a Fock state, is expected to occur in the detuning
range $|\delta|\lesssim 1$ for the chosen values of the
recoil frequency $\omega_r=0.001$ and the number
of photons $n=10$. In Fig.~\ref{fig1} we plot the evolution
of the atomic purity $P_n(\tau)$ (the left panel) and the respective power spectra (the right panel) in different
situations: (a) exact resonance $\delta=0$ and $p_0=25$, (b) a ballistic
flight with large detuning $\delta=32$ and $p_0=25$, (c)
the Doppler-Rabi resonance with $\delta=32$ and
$p_0=32000$, and (d) a chaotic walking
with $\delta=0.4$ and $p_0=25$. At exact resonance, the oscillations of purity are
frequency modulated but strictly periodic with the period
$\pi/\omega_r p_0\simeq 125.6$.
The spectrum consist of a variety of resolved peaks in the
frequency range from $\sim 0.01$ to $2.2$ in units of the vacuum Rabi frequency $\Omega_0/2\pi$
(in the right panel of Fig.~\ref{fig1}a
we show only the low-frequency and high-frequency parts
of the whole spectrum). Off-resonant oscillations of purity
is an amplitude-modulated signal the spectrum of which
contains mainly high frequencies (see Fig.~\ref{fig1}b where we
cut off for convenience the part of spectrum from $0.05$ to $5$
that does not contain any pronounced peaks). The Doppler-Rabi
oscillations and its spectrum (Fig.~\ref{fig1}c) contain the
main harmonic with $\omega/2\pi\simeq 1$ and a few high frequencies.
Purity oscillations with a randomly walking atom
look like chaotic ones with a broadened spectrum
in the frequency range from $0$ to $2.2$ (Fig.~\ref{fig1}d).
\begin{figure*}
\includegraphics[width=0.9\textwidth]{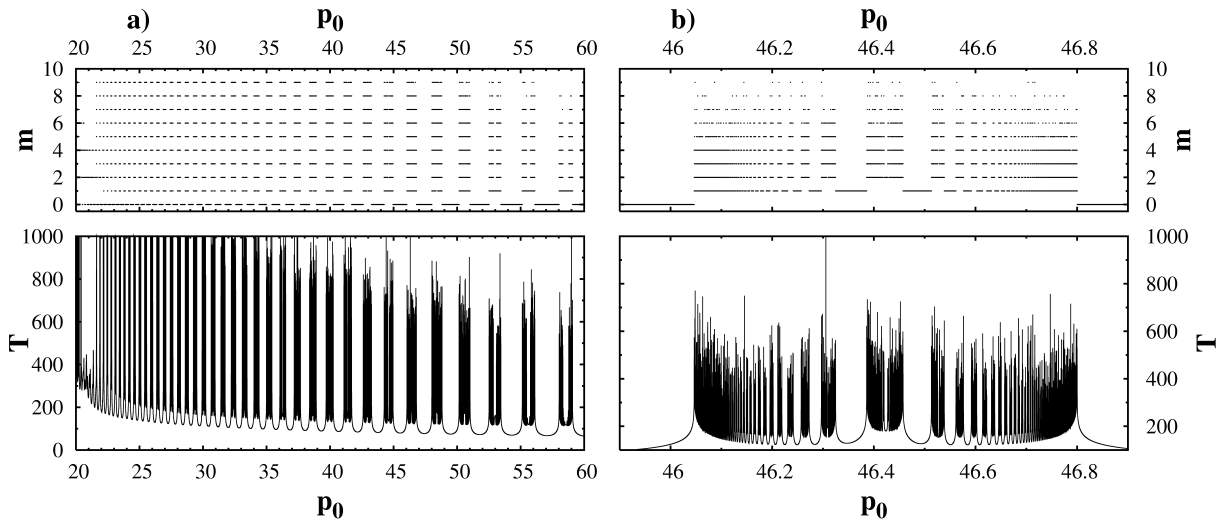}
\caption{
Fock atomic fractal (a) and a zoom of one of its fragment (b) at $n=10$, $\delta=0.4$, and
$z(0)=1$. The upper panels: how many times $m$ an
atom with a given initial momentum $p_0$ (in units of $\hbar k_f$)
changes its direction of motion before leaving a
cavity. The lower panels: the time
$T$ (in units of $\Omega_0^{-1}$) atoms with given values of
the initial momenta $p_0$ spend in the cavity before
leaving.}
\label{fig2}
\end{figure*}

The chaotic centre-of-mass walking in the quantized field has fractal 
properties similar to those that have been found with atoms in 
a classical field \cite{AP03}.
Placing atoms at the point $x=0$ with the same initial conditions and parameters but
with different values of initial momenta $p_0$, we compute the time $T(p_0)$, the atom
with a given value $p_0$ needs to reach one of the nodes of the standing wave at
$x=-\pi/2$ and $x=3\pi/2$, and the number of times $m$ when it changes its direction of motion. The
scattering function $T(p_0)$ is found to have a self-similar structure with singularities
on a Cantor-like set of initial values of momenta $p_0$. In Fig.~\ref{fig2}a we demonstrate
the mechanism of generating this set at $\delta=0.4$ in the initial Fock 
field with $n=10$ and initially excited atoms. 
The exit-time function demonstrates an intermittency of smooth
curves and complicated structures that cannot be resolved in principle, no
matter how large the magnification factor.
Fig.~\ref{fig2}b shows magnification of the function for the small interval
$45.9\leqslant p_0\leqslant 46.9$. Further magnifications reveal
a self-similar structure.

\begin{figure}
\includegraphics[width=0.4\textwidth]{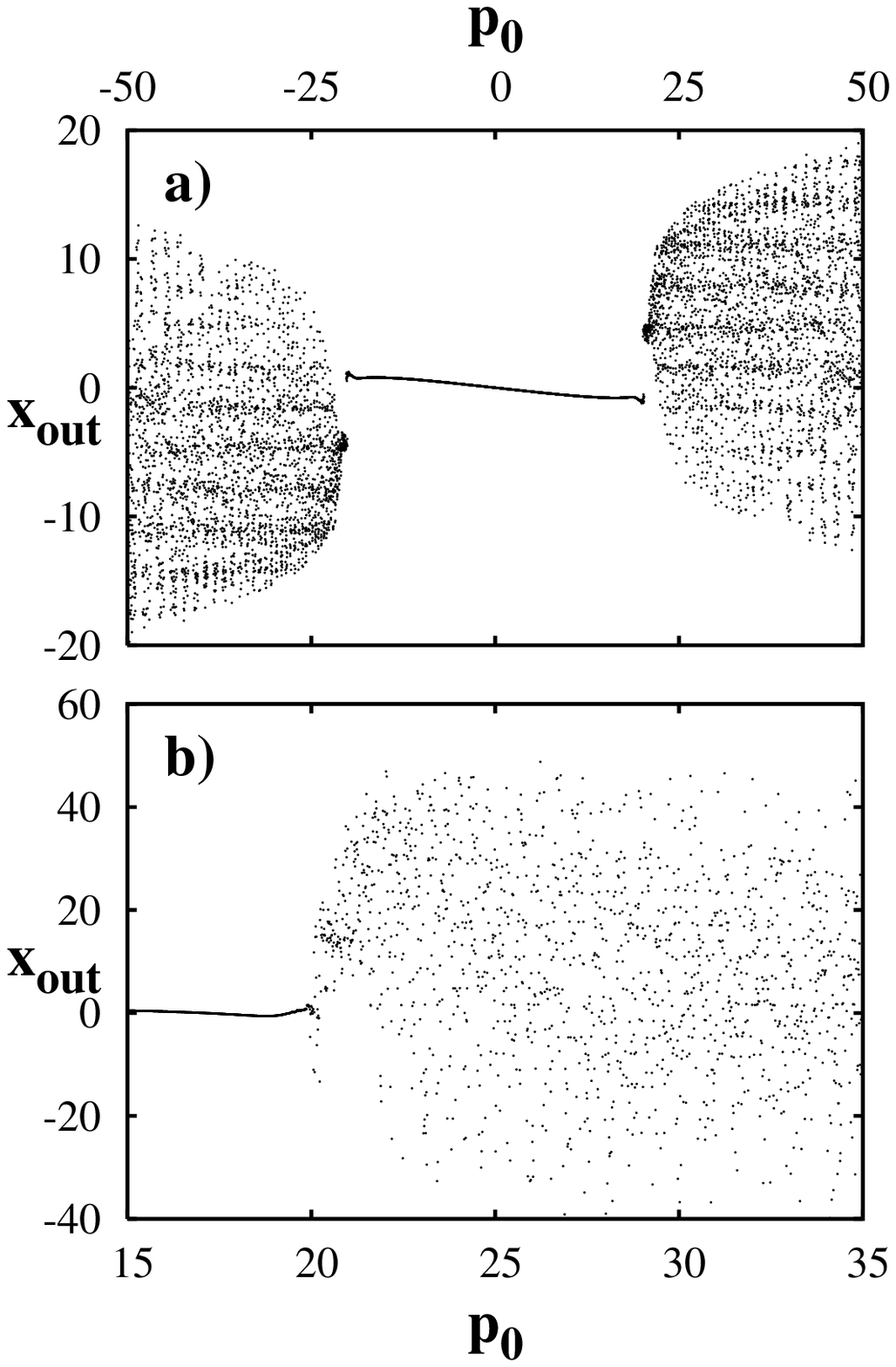}
\caption{
Sensitive dependence of the atomic position $x_\text{out}$ (in
units of $k_f^{-1}$) on the initial momentum $p_0$ at
(a) $\tau=300$ and (b) $\tau=1000$. All the other conditions
are the same as in Fig.~\ref{fig2}.}
\label{fig3}
\end{figure}
\begin{figure}
\includegraphics[width=0.4\textwidth]{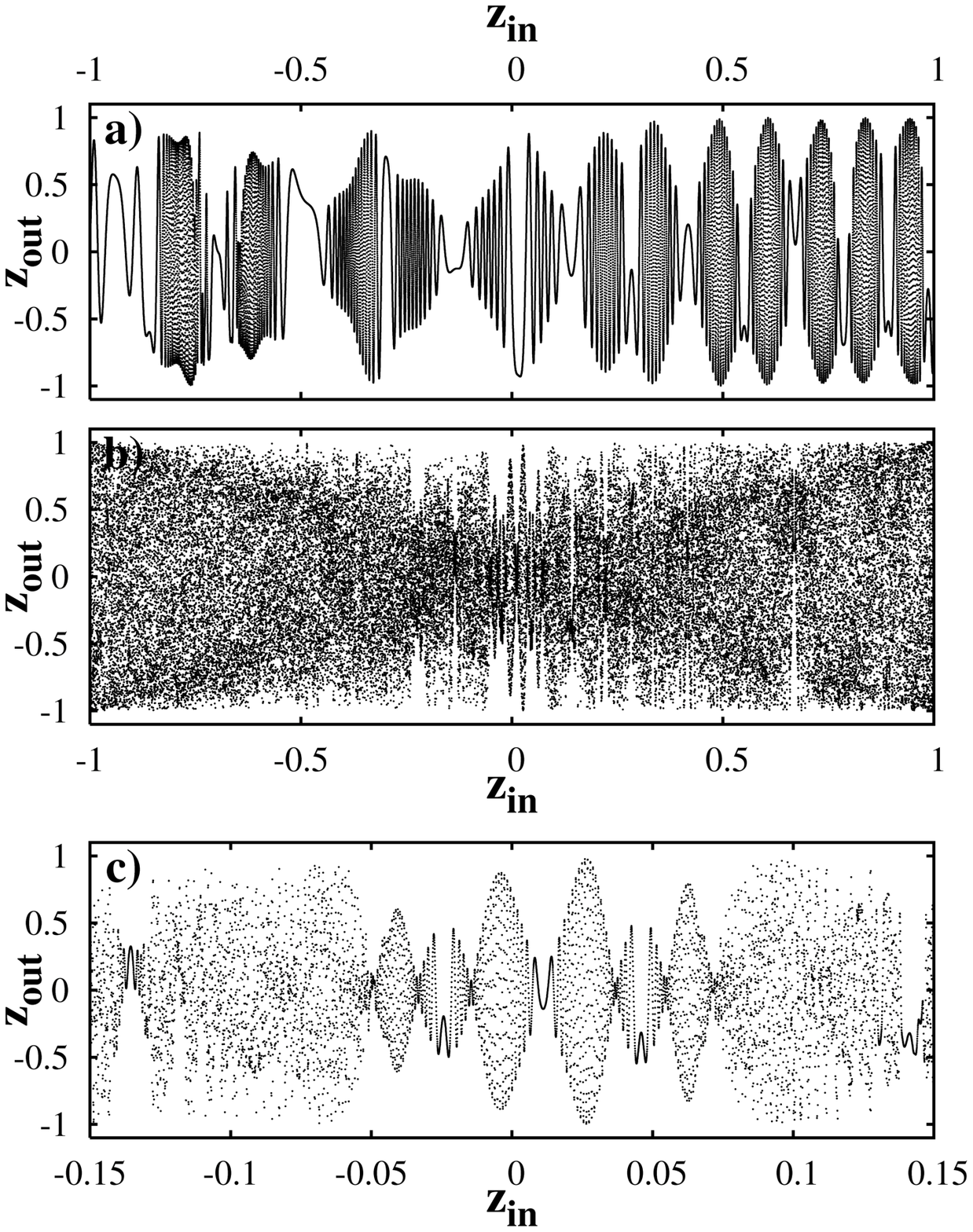}
\caption{
Sensitive dependence of the output values of the
atomic population inversion $z_\text{out}$ on its initial values $z_\text{in}$.
(a) $\tau=100$, (b) $\tau=200$, and (c) zoom of a small interval of $z_\text{in}$
around $z_\text{in}=0$ at $\tau=200$.}
\label{fig4}
\end{figure}
There are two sets of atomic 
trajectories with $T\to\infty$, the countable one consisting of separatrix-like trajectories,
corresponding to the ends of the intervals in Fig.~\ref{fig2}, and the uncountable
one consisting of trajectories with $m=\infty$. The chaotic motion can be, in principle,
verified in experiments on 1D-scattering of atoms at the standing wave. 
Fig.~\ref{fig3}
demonstrates a sensitive dependence of the atomic positions on $p_0$ 
at a fixed time moment. 
A smooth segment of this function in the range $|p_0|\lesssim 20$ 
should be attributed to atomic center-of-mass 
oscillations in the first well of the optical potential since these values of $p_0$
are not enough to overcome the respective potential barrier. When $p_0$ exceeds
a critical value, atoms leave the well, and it is practically impossible to
predict even the sign of the atomic position. The so-called predictability 
horizon can be estimated as follows:
$\tau_p\simeq \lambda^{-1}\ln{(\Delta x/\Delta x_0)}$, where $\Delta x$ is the confidence interval and
$\Delta x_0$ the  inaccuracy in
preparing initial atomic positions. In order to demonstrate the quantum-classical
correspondence in the chaotic regime we compute the dependence of the output
values of the atomic population inversion $z_\text{out}$ at a fixed moment on
its initial values $z_\text{in}$ with the other conditions and parameters 
to be the same. Fig.~\ref{fig4} shows that the predictability 
horizon of the quantum atomic degree of freedom can be estimated to be 
$\tau \simeq 200$. 

\begin{figure}
\includegraphics[width=0.4\textwidth]{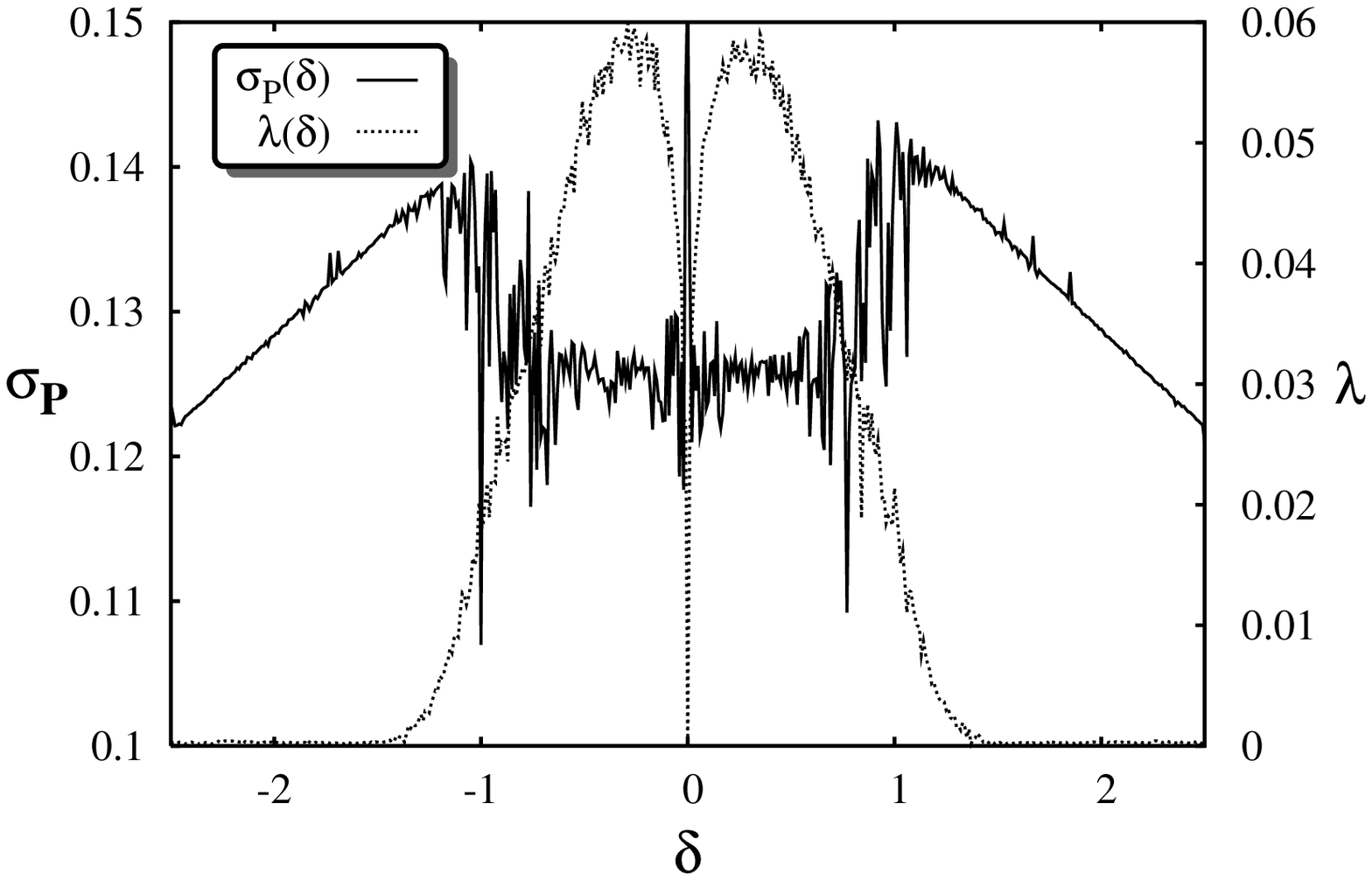}
\caption{
Quantum-classical correlation between the dependencies
of the variance of the purity $\sigma_P$ and the maximal Lyapunov
exponent $\lambda$ (in units of $\Omega_0$) on the atom-field detuning
$\delta $ (in units of $\Omega_0$). Initial Fock field with $n=10$ and an atom
prepared initially in the superposition state $(\ket{1}+\ket{2})/\sqrt{2}$.}
\label{fig5}
\end{figure}
The oscillations of purity of a chaotically moving atom and its spectrum
look like irregular ones (Fig.~\ref{fig1}d). To quantify the
irregularity we compute the root mean square
variance of purity $\sigma_P\equiv\sqrt{\aver{P^2}-\aver{P}^2}$ on a
large time scale (with $\aver{P}$ being a purity value averaged on this scale) in the range of the detuning
$|\delta|\leq 2$ and compare the result with the
dependence $\lambda(\delta)$ in the same range, where
the maximal Lyapunov exponent has been
computed with the Fock system (\ref{F}) with
the atom prepared at $\tau=0$ in the superposition state $(\ket{1}+\ket{2})/\sqrt{2}$
$(z(0)=0)$
and initial conditions $x_0=0$ and $p_0=25$. Fig.~\ref{fig5}
demonstrates that irregular oscillations of $\sigma_P$
occur on the same interval of the detuning
$\delta$, where $\lambda>0$.
Since similar correlations has been found with
different initial atomic and Fock states not only for
the linear entropy but for the von Neumann
entropy $S_N$ as well, we may
conclude that the correlation between the quantum
entanglement and the classical motion has been
established. The role of the translational atomic motion for
purity is transparent with the Fock
field when $P$ is expressed in terms of $z_n$
which are variables of two oscillators
coupled to the classical $x$--$p$ oscillator (see Eqs.~(\ref{F})). When
the latter one oscillates regularly, the first
ones do the same. When an atom walks randomly
in a cavity, it immediately causes chaotic oscillations in $z_n$ variables and therefore in purity.

\begin{figure}
\includegraphics[width=0.4\textwidth]{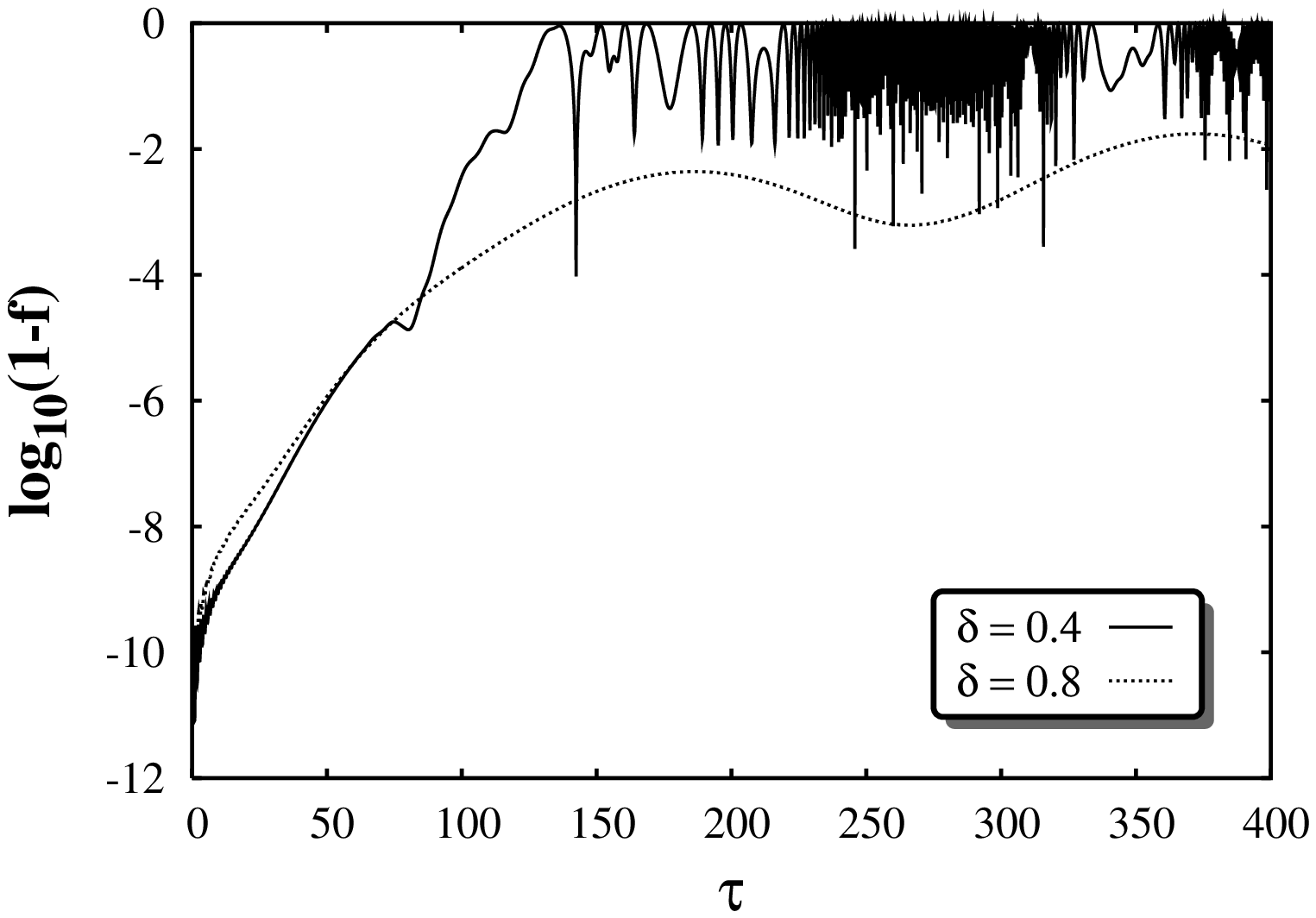}
\caption{
Time evolution of quantum fidelity (logarithmic scale)
in the chaotic (solid line at $\delta=0.4$) and regular
(dotted line at $\delta=0.8$) regimes of the center-of-mass motion
in a Fock field with $n=10$.}
\label{fig6}
\end{figure}
To quantify instability of quantum evolution of the atom-field system 
we compute the
decay of the fidelity $f(\tau)$ which is the overlap (\ref{to_fidelity}) 
of two states $\ket{\Psi_1(\tau)}$ and $\ket{\Psi_2(\tau)}$,
identical at $\tau=0$, that evolve under two Hamiltonians 
(\ref{Janes-Cum}) with the slightly different detunings $\delta$ and $\delta'\equiv\delta+\Delta\delta$.
Time evolution of fidelity with the Fock initial state $\ket{n=10}$ 
has been computed with different values of the detuning and the other 
conditions to be the same as specified above.  
For those values of $\delta$, when the center-of-mass motion is chaotic 
and the maximal Lyapunov exponent is positive, we have found that 
fidelity rapidly decays to its minimal value $f=0$ with the rate given 
approximately by the respective 
value of $\lambda$, and afterwards $f$ oscillates in an irregular manner.  
For the values of $\delta$, where the center-of-mass motion is regular  
and $\lambda \simeq 0$, fidelity evolves much more regularly.  
In Fig.~\ref{fig6} we show for convenience the evolution 
of the quantity $\log_{10}(1-f)$ in the chaotic (solid  
line, $\delta =0.4$) and regular regimes of the center-of-mass motion 
(dotted line, $\delta =0.8$) with $\Delta\delta=10^{-4}$. 
The results obtained do not depend qualitatively on the
values of differences in the detuning $\Delta\delta$.

The time evolution of fidelity can be explained as
follows. With an initially excited atom, the fidelity is
\begin{multline}
f(\tau,\,\delta,\Delta\delta)=\left|a_n^*(\tau,\delta)a_n(\tau,\delta+\Delta\delta)+
\vphantom{\sqrt{M}}\right.\\
\left.\vphantom{\sqrt{M}}
+b_{n+1}^*(\tau,\delta)b_{n+1}(\tau,\delta+\Delta\delta)\right|^2,
\end{multline}
where $a_n$ and $b_{n+1}$ are variables of a quantum oscillator coupled
to the classical $x$--$p$ oscillator. At $\delta=0.4$ and with any other values of $\delta$ when
the center-of-mass motion is chaotic, all the oscillator variables are
sensitive dependent not only to small variations of
initial conditions but to the control parameters as well. However,
the probability amplitudes and fidelity are restrictive
quantities. Because of an initial exponential divergence of the
quantities $a_n(\tau, \delta)$, $a_n(\tau, \delta+\Delta\delta)$, 
$b_{n+1}(\tau, \delta)$, and $b_{n+1}(\tau, \delta+\Delta\delta)$,
fidelity rapidly decays up to its minimal value $f=0$ with
the rate approximately given by the maximal Lyapunov
exponent of the Hamilton-Schr\"odinger equations for
a chaotically moving atom (solid line in Fig.~\ref{fig6}). After that, it oscillates
irregularly in a large range. At $\delta=0.8$ and at any other value
of the detuning when atoms move regularly in a cavity,
fidelity $f$ decays much more slowly as compared to the chaotic 
case and  oscillates smoothly (dotted line in Fig.~\ref{fig6}).

\section{\label{section_coherent}Entanglement, fidelity, and quantum-classical correlation 
in a coherent field}

In this section we consider the quantized field in a cavity to be supposed
initially in a coherent state
\begin{equation}
\ket{\alpha}=e^{-|\alpha|^2/2}\sum\limits_{n=0}^{\infty}\frac{\alpha^n}{\sqrt{n!}}\ket{n},
\label{Coh}
\end{equation}
where $|\alpha|^2=\aver{n}$
is an average number of photons in the state (\ref{Coh}).
The equations of motion (\ref{mainsys}) and (\ref{UVZSystem})
are now infinite dimensional, and their analytic
solutions have been derived in a few limiting cases in Sec.~\ref{subsection_analyt}.

Because of an infinite number of incommensurate Rabi
frequencies, the population inversion of an atom
in an initially coherent field oscillates in a complicated way.
In the Raman-Nath limit, the inversion $z(\tau)$, the atomic
purity $P(\tau)$, and the entropies $S_L$ and $S_N$
are periodic functions with 
the period $\pi/\omega_r p_0$ given by the time of the atomic
flight between two neighboring nodes of
the standing wave. Out of resonance, the center-of-mass
motion is regular at $|\delta|\gtrsim 1$ and
chaotic in approximately the same range, $|\delta|\lesssim 1$, as
in the Fock field.

\begin{figure}
\includegraphics[width=0.4\textwidth]{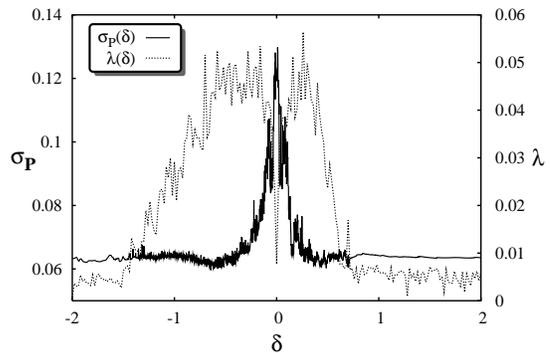}
\caption{
Quantum-classical correlation between the dependencies of the
variance of quantum purity $\sigma_P$ and the maximal Lyapunov exponent 
$\lambda$ (in units of $\Omega_0$) on the atom-field detuning $\delta$ 
(in units of $\Omega_0$). Initial coherent field with $\bar n=10$
and an atom prepared initially in the excited state.
The other conditions are the same as in Fig.~\ref{fig5}.}
\label{fig7}
\end{figure}
In numerical simulation we truncate the set 
(\ref{mainsys}) at a finite $n=100$ which is sufficient 
for all our purposes. Fig.~\ref{fig7} demonstrates the correlation between
$\delta$-dependencies of the variance of the atomic purity $\sigma_P$ and of the
maximal Lyapunov exponent $\lambda$ with the atom
prepared initially in the energetic state $\ket{2}$
and the field at $\tau=0$ in the  coherent state (\ref{Coh})
with $\bar n=10$ and the other conditions and
parameters to be the same as for the Fock
field. Comparing Figs.~\ref{fig5} and \ref{fig7}, we see that
the quantum-classical correlations in the coherent
and Fock fields are similar
despite of the fact that purity in a coherent
field even in the limiting cases oscillates in a much more complicated way
as compared to purity in a Fock field.

\begin{figure}
\includegraphics[width=0.4\textwidth]{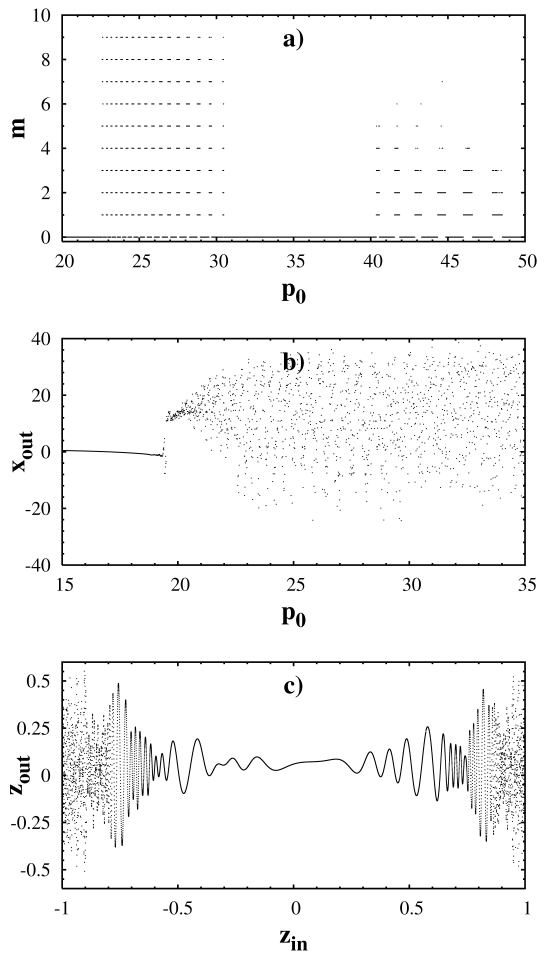}
\caption{
(a) Fractal set of the initial momenta $p_0$ (in units of $\hbar k_f$)
of atoms which 
leave a cavity with initially coherent field after $m$ turns. (b) Sensitive dependence of
the atomic position $x_\text{out}$ (in units $k_f^{-1}$) on the initial momentum $p_0$.
(c) Sensitive dependence of the output values of the atomic population
inversion $z_\text{out}$ on its initial values $z_\text{in}$. 
Control parameters $\delta=0.4$, $\bar n=10$, and $\omega_r=0.001$.}
\label{fig8}
\end{figure}
In the regime of chaotic walking, atoms in a
coherent field demonstrate fractal scattering
(Fig.~\ref{fig8}a) and sensitive dependencies on the initial
states both in the classical (Fig.~\ref{fig8}b) and
quantum (Fig.~\ref{fig8}c) degrees of freedom. The
strength of chaos, measured by the value of $\lambda$,
is the same for both the initial field states $\ket{n}$
and $\ket{\alpha}$ (see Figs.~\ref{fig5} and \ref{fig7}). The function $z_\text{out}(z_\text{in})$
is regular in a wider range of $z_\text{in}$ with atoms in a coherent field
(Fig.~\ref{fig8}c) than with atoms in a Fock field
(Fig.~\ref{fig4}) simply because of choosing a specific value of
the phase of the coherent state $\ket{\alpha}$.

\begin{figure}
\includegraphics[width=0.4\textwidth]{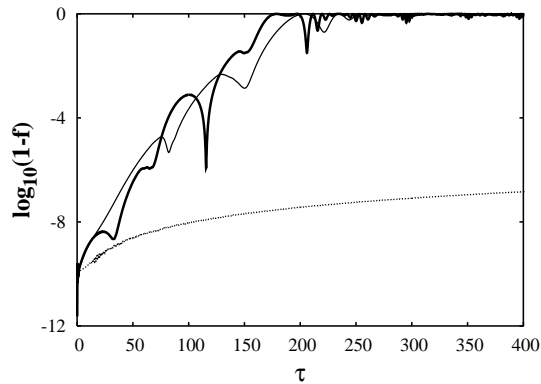}
\caption{
Time evolution of quantum fidelity (logarithmic scale) in the
chaotic (thick and thin lines) and regular (dotted line) regimes 
of the center-of-mass
motion in a coherent field with $\bar n=10$.}
\label{fig9}
\end{figure}
We have found  (see Fig.~\ref{fig8}) that with the initial momentum 
$p_0=25$ and $\delta=0.4$ the
type of atomic motion depends strongly on the initial atomic inversion $z(0)$.
If an atom is prepared initially in one of its energetic states, i.~e. $z(0)=\pm 1$, its
classical and quantum dynamics are unstable, whereas they are stable with $z(0)=0$
under the same other conditions. In Fig.~\ref{fig9} we show 
the evolution 
of the quantity $\log_{10}(1-f)$ in the regimes of chaotic walking (thick 
and thin lines,
$z(0)=\pm 1$) and regular motion (dotted line, $z(0)=0$) with $\Delta\delta
=10^{-4}$. In the chaotic regime the fidelity initially
decays exponentially with the rate $\lambda\simeq 0.04$ to be equal to the 
maximal Lyapunov exponent
computed with the set (\ref{mainsys}). This result does not depend on the 
values of
differences in the detuning $\Delta\delta$.
The fidelity practically does not 
decay in the regular regime at $z(0)=0$, and the respective maximal 
Lyapunov exponent was computed to be zero.

With randomly walking atoms,
initial decay of fidelity is practically the
same both in the Fock (Fig.~\ref{fig6}) and coherent (Fig.~\ref{fig9})
initial field states and is quantified
by the respective Lyapunov exponents.
After reaching zero value, fidelity demonstrates
erratical oscillations in
both the cases. These oscillations are more
pronounced with the Fock field
because entanglement there
occurs between a few quantum states,
whereas entanglement with a coherent
field implies, in principle, an infinite number of
states. In any case, behavior
of fidelity strongly differs with
regularly and randomly walking atoms.

\section{\label{section_concl}Conclusion}

We have found the quantum-classical correspondence in the basic
model in cavity QED by proving that entanglement between
electronic and photonic degrees of freedom and fidelity
of the Jaynes-Cummings dynamics correlate with
the center-of-mass motion of a two-level atom in
a quantized standing-wave cavity field. It has
been shown analytically and numerically both with
initial Fock and coherent field states that quantum
entropy, purity, and fidelity of regularly moving
atoms evolve in a regular way, whereas the respective
quantum evolution is unstable with atoms moving
chaotically in a periodic standing wave. Instability of the quantum
evolution has been shown to be quantified by the
respective classical maximal Lyapunov exponent for 
different initial electronic and field states and
with different values of the main control parameter,
the atom-field detuning $\delta$. We emphasize that
this quantum instability and irreversibility is
caused by internal dynamical chaos and takes place
without any  external environment. 

We have done some numerical experiments and found various manifestations
of the quantum-classical correspondence, including
dynamical chaos and fractals, which can be, in
principle, observed in real experiments with atoms
and photons in high-finesse cavities. In this connection, we should 
mention first of all pioneering experiments of Kimble's and Rempe's 
groups \cite{Monitor} on real-time position control of a single atom 
in a high-finesse microcavity where a photon may be periodically 
absorbed by the atom and re-emitted into the cavity many times before 
being lost outside the cavity. In the strong-coupling regime, when the coherent 
coupling between a single atom and intracavity field dominates atomic 
spontaneous emission and intracavity-field decay, the center-of-mass position 
within the cavity mode can be monitored in real time with high spatial and 
temporal evolution by detecting the light transmitted by the cavity. 
These achievements open up new possibilities in the control and continuous 
measurement of the internal and external dynamical variables of atoms.

\section{Acknowledgments}
The work was supported by the Program ``Mathematical methods in nonlinear 
dynamics'' of the Russian Academy of Sciences and by the Far Eastern Division
of the Russian Academy of Sciences.

\end{document}